\documentclass[preprint,amsmath,amssymb,prb,floatfix,superscriptaddress]{revtex4}
\usepackage[dvips]{graphicx}% Include figure files
\usepackage{dcolumn}% Align table columns on decimal point
\usepackage{bm}% bold math

\begin{document}
\title{Detection of coherent magnons via ultrafast pump-probe reflectance spectroscopy in multiferroic Ba$_{0.6}$Sr$_{1.4}$Zn$_2$Fe$_{12}$O$_{22}$}
\author{D. Talbayev}
\email[email: ]{diyar@lanl.gov}
\affiliation{Center for Integrated Nanotechnologies, MS K771, Los Alamos National Laboratory, Los Alamos, NM 87545, USA}
\author{S. A. Trugman}
\affiliation{Center for Integrated Nanotechnologies, MS K771, Los Alamos National Laboratory, Los Alamos, NM 87545, USA}
\author{A. V. Balatsky}
\affiliation{Center for Integrated Nanotechnologies, MS K771, Los Alamos National Laboratory, Los Alamos, NM 87545, USA}
\author{T. Kimura}
\affiliation{Division of Materials Physics, Graduate School of Engineering Science, Osaka University Toyonaka, Osaka 560-8531, Japan}
\author{A. J. Taylor}
\affiliation{Center for Integrated Nanotechnologies, MS K771, Los Alamos National Laboratory, Los Alamos, NM 87545, USA}
\author{R. D. Averitt}
\affiliation{Center for Integrated Nanotechnologies, MS K771, Los Alamos National Laboratory, Los Alamos, NM 87545, USA}
\affiliation{Department of Physics, Boston University, 590 Commonwealth Ave, Boston, MA 02215, USA}
\date{\today}

\newcommand{\cm}{\:\mathrm{cm}^{-1}}
\newcommand{\T}{\:\mathrm{T}}
\newcommand{\mc}{\:\mu\mathrm{m}}
\newcommand{\ve}{\varepsilon}
\newcommand{\dg}{^\mathtt{o}}
\newcommand{\bahex}{Ba$_{0.6}$Sr$_{1.4}$Zn$_2$Fe$_{12}$O$_{22}\ $}

\begin{abstract}
We report the detection of a magnetic resonance mode in multiferroic
Ba$_{0.6}$Sr$_{1.4}$Zn$_2$Fe$_{12}$O$_{22}$ using time domain pump-probe reflectance spectroscopy. Magnetic sublattice precession is coherently excited via picosecond thermal modification of the exchange energy. Importantly, this precession is recorded as a change in reflectance caused by the dynamic magnetoelectric effect. Thus, transient reflectance provides a sensitive probe of magnetization dynamics in materials with strong magnetoelectric coupling, such as multiferroics, revealing new possibilities for application in spintronics and ultrafast manipulation of magnetic moments.
\end{abstract}

\maketitle

%\renewcommand{\baselinestretch}{1.7}
%\normalsize

In magneto-electric (ME) materials, or multiferroics, it is possible to modify the magnetization with an applied electric field or $vice$ $versa$\cite{fiebig:123, eerenstein:759, cheong:13, ramesh:21}. Recent experiments have demonstrated magnetic switching of electric polarization and switching of magnetization by electric field, thereby opening remarkable opportunities in spintronics and magnetic recording \cite{kimura:55, hur:392, lottermoser:541, zavaliche:1793}. The dynamical properties of multiferroics are determined by magnetic and lattice vibrations and by the hybridization between them \cite{katsura:27203}, which can result in electric-dipole active magnons\cite{pimenov:97, sushkov:27202}. Ultrafast optical spectroscopy can provide insight into the fundamental microscopic dynamics and, in particular, the coupling between multiple degrees of freedom which determine the underlying functional response of complex materials. Thus, pump-probe studies can answer technological questions such as achievable switching speeds in multiferroics.

In a pump-probe study of multiferroic LuMnO$_3$, Lim $et$ $al$ observed the coherent excitation of optical and acoustic phonons in the material by femtosecond optical pulses\cite{lim:4800}. The associated speed of sound exhibited an anomaly at the antiferromagnetic transition temperature, which is a signature of the coupling between elasticity and magnetism. Takahashi $et$ $al$ investigated the response of multiferroic BiFeO$_3$ thin films to a  femtosecond optical excitation \cite{takahashi:117402}. They found that the injection of charge carriers and their subsequent acceleration by an applied voltage leads to an ultrafast modulation of the electric polarization, which corresponds to the coherent excitation of the over-damped soft optical phonon. The time-dependent polarization emits terahertz radiation that can be used to detect the polarization direction and image the polarization domains in the BiFeO$_3$ film. In addition to controlling the lattice dynamics, femtosecond optical pulses have been employed to manipulate and detect the magnetic state of solids \cite{vankampen:227201, kimel:43201}. Although we are not aware of any pump-probe studies of magnetization dynamics in multiferroics, tremendous progress has been made in the understanding of opto-magnetic phenomena in magnetic insulators, where various mechanisms (such as the inverse Faraday effect) of magnon excitation and detection have been elucidated \cite{kimel:43201}. The ability to probe both phonon and magnon response in the time-domain is an essential characteristic of pump-probe spectroscopy which makes it uniquely suited for investigation of the dynamical properties of multiferroics.

In this letter, we report coherent optical excitation and detection of magnons in multiferroic Ba$_{2-x}$Sr$_x$Zn$_2$Fe$_{12}$O$_{22}$ (BSZFO), a compound with a large ME response \cite{kimura:137201}. Coherent magnon excitation can be viewed as triggering magnetic sublattice precession by a femtosecond optical pulse. In BSZFO, the precession is triggered when the sublattice equilibrium position is modified through picosecond heating of the crystalline lattice from energy relaxation of an initial optically excited electron distribution. This results in a rapid modification of the exchange coupling between sublattices resulting in coherent magnon generation. We detect the precessing magnetic moment as an oscillation of the reflectance. This sharply contrasts with conventional detection schemes that employ Kerr or Faraday rotation \cite{vankampen:227201, kimel:43201}. The modulation of the reflectance results from a strong ME response, i.e., the dependence of the dielectric constant  $\epsilon(\lambda)$ at optical wavelengths $\lambda$ on the magnetic state that is modified by the precessing magnetization. This cross-coupling between a magnetic precession mode and an electronic excitation is a manifestation of the ME coupling, a dynamic ME effect. A similar consequence of the dynamic ME effect is the observation of electromagnons, the mixed resonant magnetic-lattice vibrations\cite{pimenov:97, sushkov:27202}.

Single crystals of BSZFO with $x$=1.4 were grown from Na$_2$O-Fe$_2$O$_3$ flux following Momozawa $et$ $al$\cite{momozawa:403}. The setup for ultrafast measurements uses a 250-kHz repetition rate Ti:Sapphire regenerative amplifier producing 6-$\mu$J, 80-fs pulses at 800 nm. The amplifier output is converted to 400-nm pump pulses by second harmonic generation. The pump fluence is in the 100-500 $\mu$J/cm$^2$ range. The onset of electronic absorption in BSZFO is around 600 nm as determined by room temperature ellipsometry. This dictates the choice of the pump wavelength, since the negligible absorption at long wavelengths prevents the detection of pump-induced reflectance changes. We use probe pulses with wavelengths of 400 and 800 nm in this study. The probe power is always $\leq 5\%$ of the pump power. In the pump-probe experiment, the strong pump pulse excites the material, while the delayed probe pulse measures the pump-induced change in the material's reflectance. The pump-probe measurements are conducted at near-normal incidence on the (001) face of the BSZFO sample mounted on the cold-finger of a helium-flow cryostat in the center of a split-coil superconducting magnet.
\begin{figure}[ht]
\begin{center}
\includegraphics[width=3in]{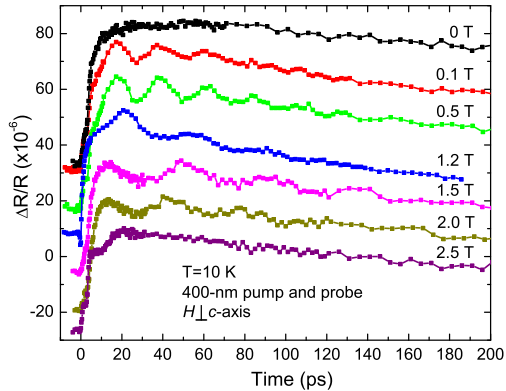}
\caption{\label{fig:dror}(Color online) Pump-probe reflectance of BSZFO in magnetic field. The spectra are shifted along the vertical axis for clarity.}
\end{center}
\end{figure}

BSZFO crystallizes in the hexagonal structure of the space group
$R\bar{3}m$ ($Z=3$), and shows magnetic ordering below $T_{\rm
N}\approx$ 325 K. The magnetic ground state exhibits a helimagnetic
structure with the propagation vector along the $c$-axis \cite{kimura:137201,momozawa:771,momozawa:1292}. A magnetic field applied perpendicular to the $c$-axis induces a series of magnetic phases ranging from a slightly modified helimagnet to a collinear ferrimagnet \cite{momozawa:771,momozawa:1292}. BSZFO exhibits magnetic-field-induced electric polarization switching that is compatible with the variation of magnetic structures\cite{kimura:137201}. 

Figure~\ref{fig:dror} shows the pump-probe reflectance spectra at 10 K, well below $T_N$. In zero magnetic field, the spectrum consists of a $\sim 10$-ps rise followed by a slowly-decaying plateau attributed to the initial electronic transfer between the 3d states of neighboring Fe ions \cite{knizek:153103} and the subsequent thermalization of
electronic, lattice, and spin systems. Upon application of a field of 0.1 T, a distinct oscillatory response appears. Further increase of the field leads to an increase in the oscillation period. At 2.5 T, in the collinear ferrimagnetic phase, the oscillatory response disappears. The frequency of the reflectance oscillation (RO) does not change when the sample is electrically poled following the procedure described by Kimura $et$ $al$\cite{kimura:137201}.
\begin{figure}[ht]
\begin{center}
\includegraphics[width=3in]{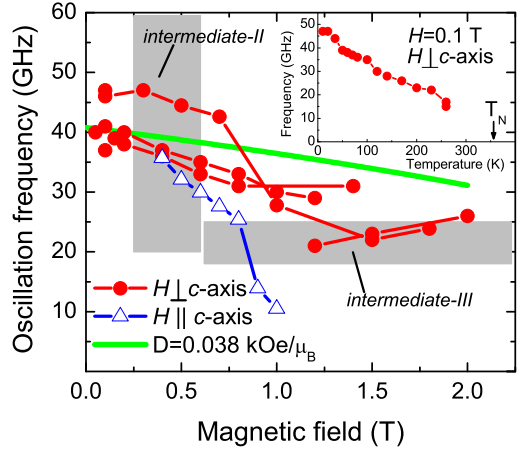}
\caption{\label{fig:freqfield}(Color online) The frequency of the RO at $T=10$ K as a function of magnetic field. Solid line - resonance frequencies calculated from Eq. (\ref{eq:freen}). The upper inset shows the temperature dependence of the RO frequency.}
\end{center}
\end{figure}

\begin{figure}[ht]
\begin{center}
\includegraphics[width=3in]{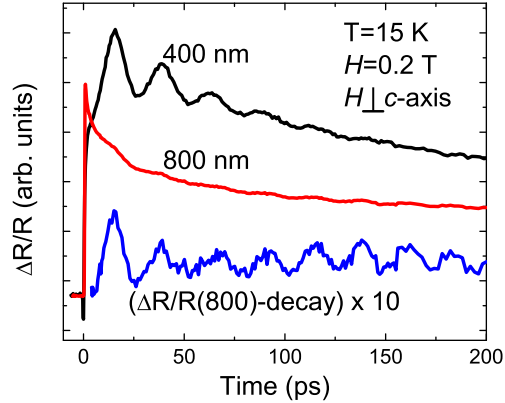}
\caption{\label{fig:wavedep}(Color online) Pump-probe reflectance at 400 nm and 800 nm probe wavelengths. The lowest curve is the reflectance at 800 nm after subtraction of the decay background multiplied by 10. The RO frequencies are: f(400) = 39.7 GHz, f(800) = 40.9 GHz.}
\end{center}
\end{figure}

Figure \ref{fig:freqfield} shows the field dependence of the RO frequency. The near zero-field frequency of $\sim40$ GHz and its pronounced magnetic-field dependence strongly suggest a magnetic origin of the RO, i.e., the coherent excitation of a $k$=0 magnon in BSZFO. This conclusion is supported by the temperature dependence of the RO frequency measured at 0.1 T (the inset of Fig. \ref{fig:freqfield}). The RO could not be detected at temperatures above 260 K. The RO frequency tends to zero in the vicinity of the N\'eel temperature and qualitatively follows the temperature dependence of the magnetic order parameter\cite{momozawa:1292}, in agreement with the behavior of the $k$=0 magnon frequency in antiferromagnetic and ferrimagnetic systems\cite{foner:383}. The RO cannot result from a coherent lattice response, as the measured frequency is too low to be explained by coherent optical phonons excited by stimulated Raman scattering \cite{merlin:207}. We experimentally rule out coherent acoustic phonons because the RO oscillation frequency does not depend on the probe wavelength (Fig.~\ref{fig:wavedep}). The acoustic phonon response would lead to a $\propto n/\lambda$ dependence of the RO frequency, where $n$ is the refractive index of the crystal at the wavelength $\lambda$\cite{thomsen:4129,wu:85210}. The same RO frequency at 400 and 800 nm would imply that $2\,n$(800)=$n$(400) at 15 K. Our room temperature ellipsometry yields $n$(400)=2.64 and $n$(800)=2.44.

The measured RO results from the modulation of the refractive index $n$ by the motion of Fe spins. The dynamic modulation of $n$ by the precessing magnetization is similar to the modulation of $n$ by a static magnetic field found in several complex materials \cite{rai:235101,rai:75112}. Although recent $ab$ $initio$ calculation of the electronic properties of BSZFO\cite{knizek:153103} neglects the effects of spin orientation, the electronic structure of Ni$_3$V$_2$O$_8$ shows a quantitative dependence of the optical conductivity on the spin orientation\cite{rai:235101}.

Precessing sublattice magnetization can modify not only the diagonal components of the dielectric tensor, but also the off-diagonal components $\kappa$ responsible for the magneto-optical Kerr effect (MOKE). In our experiment, the measured probe intensity is proportional to $|r_{ss}|^2$ and $|r_{pp}|^2$ for $s$- and $p$-polarized incident probe; $r_{ss}$ and $r_{pp}$ are Fresnel reflection coefficients. Only $r_{pp}$ depends on $\kappa$ in transverse MOKE\cite{you:541}. However, the reflectance spectra measured with $s$- and $p$-polarized probe display identical RO, which demonstrates that the RO cannot result from $\kappa$-dependence of $r_{pp}$. We conclude that the RO arises from the modification of the diagonal components of the dielectric tensor by the magnetization motion, the dynamic ME effect.

\begin{figure}[ht]
\begin{center}
\includegraphics[width=3in]{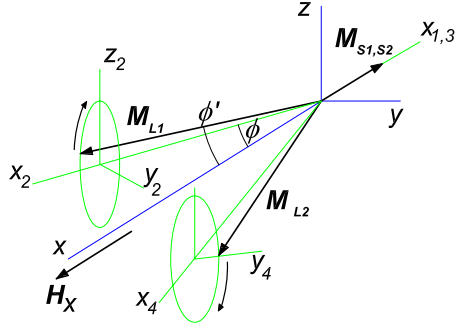}
\caption{\label{fig:mageigstate}(Color online) Motion of magnetization vectors in the rocking magnon mode in the intermediate-II phase. Pump-induced modification of the exchange energy changes the equilibrium angle of the vector $\bm{M_{L1}}$ from $\phi'$ to $\phi$, while the vectors $\bm{M_{S1}}$ and $\bm{M_{S2}}$ remain in equilibrium along the negative $x$-direction. The vectors $\bm{M_{L1}}$ and $\bm{M_{L2}}$ start precessing along the ellipses as indicated by arrows. The axes $x_1,x_2,x_3$, and $x_4$ show the equilibrium directions of the corresponding sublattices.}
\end{center}
\end{figure}

Next, we present a theoretical description of the magnetization precession that further supports the magnetic origin of the RO. In zero field, BSZFO adopts the helical magnetic structure consisting of two kinds of alternating ferrimagnetic blocks ($M_S$ and $M_L$) with antiferromagnetic exchange interaction between nearest and next-nearest neighbor blocks - $J_{LS}$, $J_{LL}$, and $J_{SS}$\cite{momozawa:1292}. We numerically solve the equations of motion of magnetization in the intermediate-II phase of BSZFO with static magnetic field applied perpendicular to the $c$-axis. The intermediate-II phase is commensurate and consists of two small ($M_{S1}$ and $M_{S2}$) and two large ($M_{L1}$ and $M_{L2}$) spin blocks (Fig.~\ref{fig:mageigstate}). The exchange constants $J_{LS}=51.2$ kOe/$\mu_B$, $J_{LL}=2.52$ kOe/$\mu_B$, and $J_{SS}=6.52$ kOe/$\mu_B$ for this phase were deduced from magnetization measurements\cite{momozawa:1292} assuming that the magnetizations $M_S$ and $M_L$ are confined to the $c$-plane by easy-plane anisotropy. To describe this anisotropy, we include single ion terms of the form $DM_z^2$ in the magnetic free energy:
\begin{eqnarray}
F & = & J_{LS}\sum_{i,j=1,2}(\bm{M_{Si}}\cdot\bm{M_{Lj}})
+2J_{LL}\bm{M_{L1}}\cdot\bm{M_{L2}}\nonumber\\
  &   & +2J_{SS}\bm{M_{S1}}\cdot\bm{M_{S2}} + D\sum_{i=1,2}(M_{Liz}^2+M_{Siz}^2)\nonumber\\
  &   &  - H\sum_{i=1,2}(M_{Lix}+M_{Six}),
	\label{eq:freen}
\end{eqnarray}
where the $z$-direction is along the $c$-axis and the static field $H$ is along the $x$-axis. We solve for the $k$=0 normal modes and resonance frequencies of the magnetic system described by equation (\ref{eq:freen}) and find that the excitation spectrum consists of two low- and two high-frequency modes. When D=0, the frequency of the lowest mode is zero at all fields and corresponds to a global rotation about the direction of the applied static field. The other low-frequency mode corresponds to the Larmor precession of the total magnetic moment about the applied magnetic field. When $D>0$, the former mode acquires a finite frequency that decreases with increasing field, as illustrated by the solid line in Fig.~\ref{fig:freqfield}. The corresponding magnon eigenstate describes the rocking motion of the plane spanned by the vectors $\bm{M_{L1}}$ and $\bm{M_{L2}}$ about the direction of the applied field (Fig.~\ref{fig:mageigstate}). The motion of magnetization vectors $\bm{M_{S1}}$ and $\bm{M_{S2}}$ is nil in this eigenstate.

The rocking magnon mode frequency decreases with increasing field, similar to the measured RO frequency, which lends further support to the interpretation of the RO as the modulation of the index $n$ by the coherent magnon. We fit the measured resonance frequencies in the 0.25-0.6 T field range (shaded area in Fig.~\ref{fig:freqfield} corresponding to the intermediate-II phase) and find $D=0.038$ kOe/$\mu_B$, the only free parameter in our calculation. Fig.~\ref{fig:freqfield} also shows the resonance frequency calculation extrapolated into the other magnetic phases. The calculation reproduces well the qualitative magnetic field behaviour of the measured frequency in the intermediate-III phase that consists of the same spin blocks as the intermediate-II phase but is separated from it by a discontinuous drop in the angle $\phi$ (Fig.~\ref{fig:mageigstate}).

To excite the rocking magnon mode, the absorbed pump energy modifies the exchange interaction between magnetic sublattices as the electron and phonon systems thermalize on the time scale of several picoseconds. (The temperature dependence of the couplings $J_{LS}$, $J_{LL}$, and $J_{SS}$ was measured by Momozawa and Yamaguchi\cite{momozawa:1292}.) The thermally modified exchange couplings lead to a new equilibrium angle $\phi$ for the large spin blocks $\bm{M_{L1}}$ and $\bm{M_{L2}}$, while the equilibrium of the small blocks $\bm{M_{S1}}$ and $\bm{M_{S2}}$ remains undisturbed (Fig.~\ref{fig:mageigstate}). The off-equilibrium sublattices start a coherent motion that corresponds precisely to the rocking mode since it is the only mode in which the vectors $\bm{M_{L1}}$ and $\bm{M_{L2}}$ are displaced symmetrically with respect to the $x$-axis. No other magnons can be excited by this process.

The described model explains why the RO is absent from the high-field (above 2.5 T) and zero-field spectra. Above 2.5 T, BSZFO enters a collinear ferrimagnetic phase in which the equilibrium of the sublattices is unaltered by small variations of the exchange couplings. In zero field, BSZFO adopts the helical phase and the equilibrium positions of the sublattices may change with varying exchange couplings. Thus, photo-excitation of magnons seems possible. However, the magnon eigenstates, their excitation efficiency and coupling to the dielectric tensor differ completely from the above picture. The combination of these factors results in the absence of the RO in the zero field spectra.

To summarize, we demonstrate the dynamic ME effect in multiferroic BSZFO, a modulation of the material's dielectric tensor by magnetization precession. This effect allows the detection of coherent magnons via pump-probe reflectance, which is conceptually and technically simpler than the detection via polarization rotation and enables potential applications of the phenomenon in spintronics and ultrafast manipulation of the magnetic state. (Similar phenomenon was observed via birefringence in the antiferromagnet TmFeO$_3$\cite{kimel:850}.) As the modulation of dielectric constants by magnons results from the ME coupling, multiferroic materials with strong ME response present a fertile ground for further investigations.

We thank Elbert Chia and Andrew Dattelbaum for their help with reflectance and ellipsometry measurements. This work was supported by the LDRD program at Los Alamos National Laboratory.

%\newpage
%\bibliography{hexaferrite}

\end{document}